\documentclass[12pt]{article}
\usepackage{blindtext}
\usepackage{scrextend}
\usepackage{hyperref}
\usepackage{amsmath}
\usepackage{amsfonts}
\usepackage{amssymb}
\usepackage{amsthm}                                                                                                                                                                                                                                        
\usepackage{bm}
\usepackage{times}
\usepackage{cancel}
\usepackage{lscape}
\usepackage{graphicx}
\usepackage[super]{natbib}
\usepackage{leftidx}    
\graphicspath{ {images/} }
\usepackage[tikz]{bclogo}
\usetikzlibrary{arrows}
\usetikzlibrary{decorations.pathreplacing}
\usepackage{booktabs}
\usepackage{multirow}
\usepackage{pgf,tikz}
\tikzstyle{every node}=[font=\normalsize]
\usepackage{color}
 \usepackage{relsize}
\usepackage{soul}
\usepackage{verbatim}

\newcommand{\ket}[1]{| #1 \rangle}

\title{Why Bohr Was (Mostly) Right}
\author{Jeffrey Bub\\ \small Philosophy Department\\\small Institute for Physical Science and Technology\\\small Joint Center for Quantum Information and Computer Science\\  \small University of Maryland, College Park, MD 20742, USA}

\date{}
\normalsize
\begin{document}

\maketitle

\begin{abstract}
After a discussion of the Frauchiger-Renner argument that no ``single-world'' interpretation of quantum mechanics can be self-consistent, I propose a ``Bohrian'' alternative to many-worlds or QBism as the rational option.
\end{abstract}

In a January 17, 2013 post to his blog \emph{Preposterous Universe},\cite{Carroll} Sean Carroll commented on a poll about foundational attitudes to quantum mechanics.\cite{SKZ} The title of Carroll's blog post was ``The Most Embarrassing Graph in Modern Physics.'' One of the poll questions was ``What is your favorite interpretation of quantum mechanics?'' The responses were mixed, with a plurality of 42\% for the Copenhagen interpretation. Carroll comments:
\begin{quote}
I'll go out on a limb to suggest that the results of this poll should be very embarrassing to physicists. Not, I hasten to add, because Copenhagen came in first, although that's also a perspective I might want to defend (I think Copenhagen is completely ill-defined, and shouldn't be the favorite anything of any thoughtful person). The embarrassing thing is that we don't have agreement.
\end{quote}
The Copenhagen interpretation, a motley collection of commentary by physicists like Heisenberg, Pauli, Rosenfeld, and others in the Bohr camp, is not quite the same as Bohr's complementarity interpretation, as the response to the question in the poll about Bohr's view of quantum mechanics attests---only 21\% thought it was correct and 27\% thought it was wrong. 

I want to suggest a way of understanding Bohr that is not at all ill-defined and \emph{should} be the favorite interpretation of any thoughtful person. To be clear, I don't intend to offer a scholarly exegesis of Bohr's essays on quantum mechanics or to quote Bohr to support a claim about what he \emph{really} meant. Rather, I will argue that this is what Bohr ought to have meant, and that what he ought to have meant is not so far off from what he wrote on complementarity. 

To motivate the discussion, I first want to consider a recent ``no go'' result by Frauchiger and Renner,\cite{FrauchigerRenner} a novel re-formulation of the ``Wigner's friend'' argument,\cite{Wigner} with a twist that exploits Hardy's paradox.\cite{Hardy} The argument purports to show that no ``single-world'' interpretation of quantum mechanics can be self-consistent, where a single-world interpretation is any interpretation that asserts that, in a measurement with multiple possible outcomes, just one outcome  actually occurs. Frauchiger and Renner conclude that ``we are forced to give up the view that there is one single reality.''

Here's how the argument goes:  Alice measures an observable $A$ with eigenstates $\ket{h}_{A}, \ket{t}_{A}$ on a system in the state $\frac{1}{\sqrt{3}}\ket{h}_{A} + \frac{\sqrt{2}}{\sqrt{3}}\ket{t}_{A}$. One could say that Alice ``tosses a biased quantum coin'' with probabilities 1/3 for heads and 2/3 for tails. She prepares a qubit in the state $\ket{0}_{B}$ if the outcome is $h$, or in the state  $\frac{1}{\sqrt{2}}(\ket{0}_{B} + \ket{1}_{B})$ if the outcome is $t$, and sends it to Bob. When Bob receives the qubit, he measures the qubit observable $B$ with eigenstates $\ket{0}_{B}, \ket{1}_{B}$. After Alice and Bob obtain definite outcomes for their measurements, the quantum state of the combined quantum coin and qubit system is $\ket{h}_{A}\ket{0}_{B}$ or $\ket{t}_{A}\ket{0}_{B}$ or  $\ket{t}_{A}\ket{1}_{B}$, with equal probability.  At least, that's the state of quantum coin and qubit system from the perspective of Alice and Bob.
 
Now, the quantum coin and the qubit, as well as Alice and Bob, their measuring instruments and all the systems in their laboratories that become entangled with the measuring instruments in registering and recording the outcomes of the quantum coin toss and the qubit measurement, including the entangled environment, are just two big many-body quantum systems $S_{A}$ and $S_{B}$, which are assumed to be completely isolated from each other after Bob receives Alice's qubit The symbols $A$ and $B$ could represent ``super-observables'' of  the composite systems $S_{A}$ and $S_{B}$ that  end up with definite values corresponding to the outcomes of Alice's and Bob's measurements on the quantum coin and the qubit.

Consider a super-observer, with vast technological abilities, who contemplates measuring a super-observable $X$ of $S_{A}$ with eigenstates $\ket{\mbox{fail}}_{A} =  \frac{1}{\sqrt{2}}(\ket{h}_{A} + \ket{t}_{A}), \ket{\mbox{ok}}_{A} = \frac{1}{\sqrt{2}}(\ket{h}_{A} - \ket{t}_{A})$, and a super-observable  $Y$ of $S_{B}$ with eigenstates $\ket{\mbox{fail}}_{B} =  \frac{1}{\sqrt{2}}(\ket{0}_{B} + \ket{1}_{B}),\ket{\mbox{ok}}_{B} =  \frac{1}{\sqrt{2}}(\ket{0}_{B} - \ket{1}_{B})$, where $\ket{h}_{A},\ket{t}_{A}$ and $\ket{0}_{B}, \ket{1}_{B}$ now represent eigenstates of the super-observables $A$ and $B$, and $\{h,t\}$ and $\{0,1\}$ represent the corresponding eigenvalues.\footnote{Since  $S_{A}$ and $S_{B}$ are  isolated systems, we should really consider two super-observers: a super-observer who measures $X$ on $S_{A}$ and a second super-observer who measures $Y$ on $S_{B}$. For simplicity, I'll continue to refer to a super-observer who measures $X\otimes Y$, bearing in mind that this  is really a composite of two super-observers.}

Of course, such a measurement would be extraordinarily difficult to carry out in practice on the whole composite system, including all the systems in the environment, but nothing in quantum mechanics precludes this possibility. From the super-observer's perspective,  $S_{A}$ and $S_{B}$ are just composite many-body entangled quantum systems  that have evolved unitarily to a combined entangled state:
\begin{equation}
\ket{\psi}  =  \frac{1}{\sqrt{3}}(\ket{h}_{A}\ket{0}_{B} + \ket{t}_{A}\ket{0}_{B} +\ket{t}_{A}\ket{1}_{B}) 
\end{equation}
The $A$ and $B$ values simply don't appear anywhere in the super-observer's description of events, so the super-observer sees no reason to conditionalize the state to one of the product states $\ket{h}_{A}\ket{0}_{B}$ or $\ket{t}_{A}\ket{0}_{B}$ or  $\ket{t}_{1}\ket{1}_{B}$. For the super-observer, this would seem to require a suspension of unitary evolution in favor of an unexplained collapse of the quantum state.

But now we have a contradiction. The state $\ket{\psi}$ can also be expressed as:
\begin{eqnarray}
\ket{\psi} & = & \frac{1}{\sqrt{12}}\ket{\mbox{ok}}_{A}\ket{\mbox{ok}}_{B} - \frac{1}{\sqrt{12}}\ket{\mbox{ok}}_{A}\ket{\mbox{fail}}_{B} \nonumber\\
&& + \frac{1}{\sqrt{12}}\ket{\mbox{fail}}_{A}\ket{\mbox{ok}}_{B} + \sqrt{\frac{3}{4}}\ket{\mbox{fail}}_{A}\ket{\mbox{fail}}_{B}\\
& = & \sqrt{\frac{2}{3}}\ket{\mbox{fail}}_{A}\ket{0}_{B} + \frac{1}{\sqrt{3}}\ket{t}_{A}\ket{1}_{B}\\
& = & \frac{1}{\sqrt{3}}\ket{h}_{A}\ket{0}_{B} + \sqrt{\frac{2}{3}}\ket{t}_{A}\ket{\mbox{fail}}_{B}
\end{eqnarray}
From the first expression for $\ket{\psi}$, the probability is $1/12$ that the super-observer finds the pair of outcomes $\{\mbox{ok, ok}\}$ in a joint measurement of  $X$ and $Y$ on the two systems. But this outcome is \emph{inconsistent with any pair of outcomes for Alice's and Bob's measurements}. From the second expression, the pair $\{\mbox{ok}, 0\}$ has zero probability, so $\{\mbox{ok}, 1\}$ is the only possible pair of values for the super-observables $X, B$ if $X$ has the value $\mbox{ok}$. From the third expression, the pair $\{t, \mbox{ok}\}$ has zero probability, so $\{h, \mbox{ok}\}$ is the only possible pair of values for the super-observables $A, Y$ if $Y$ has the value $\mbox{ok}$. But the pair of values $\{h, 1\}$ for the super-observables $A$ and $B$ has zero probability in the state $\ket{\psi}$, so it does not correspond to a possible pair of measurement outcomes for Alice and Bob.\footnote{Since the composite systems $S_{A}$ and $S_{B}$ are assumed to be completely isolated from each other, the no-signaling principle ensures that the outcome of a measurement on one of the systems can't depend on whether or not a measurement is performed on the other system.}

Both the observer (Alice and Bob), and the super-observer (the two super-observers who measure $S_{A}$ and $S_{B}$) apply quantum mechanics correctly. The argument depends only on (i) the one-world assumption, that a measurement has a single outcome, (ii)  the assumption that quantum mechanics applies to systems of any complexity, including observers, and (iii) self-consistency, in particular agreement between an observer and a super-observer. The surprising conclusion is that there is no consistent story that includes an observer and a super-observer: a possible outcome, according to quantum mechanics, of a super-observer's measurement of a super-observable $X\otimes Y$ of the whole composite observer system is inconsistent with the observer obtaining an actual outcome for any of the possible values of the measured observable $A\otimes B$. It's the theoretical possibility of a super-observer that shows the inconsistency of the theory. As far as we know, there are no super-observers, but the actuality of a measurement outcome can't depend on whether or not a super-observer turns up at some point.

What are the options? David Wallace makes a useful distinction between  \emph{representational} and \emph{probabilistic} interpretations of the quantum state.\cite{Wallace} If the quantum state is interpreted representationally, as the analogue of the classical state in stipulating what's true and what's false, and we accept assumption (ii) and hence the universality of unitarity (so no collapse of the quantum state), the correct description of the composite system $S_{A} + S_{B}$ just before the super-observer's measurement is the entangled state (2), a superposition with several components, each associated with a different measurement outcome. Dropping assumption (i) then leads to Everett's many-world interpretation, the preferred option of Frauchiger and Renner. 

If we interpret the quantum state probabilistically, then we seem to be forced to QBism, the quantum Bayesianism of Christopher Fuchs and Ruediger Schack.\cite{Fuchs1} The QBist rejects assumption (iii), the self-consistency assumption. On this view, all probabilities, including quantum probabilities, are understood in the subjective sense as the personal judgements of an agent, based on how the external world responds to  actions by the agent. For QBists, the Born rule ``is a normative statement \ldots  about the decision-making behavior any individual agent should strive for \ldots not a ``law of nature'' in the usual sense,'' and ``measurement outcomes \emph{just are} personal experiences for the agent gambling upon them.''\cite{Fuchs2} So there is no requirement that the perspective of an observer and a super-observer should be consistent.

There is another option if we interpret the quantum state probabilistically, which is to reject assumption (ii)---not by restricting the universality of the unitary dynamics or any part of quantum mechanics, but by placing a constraint on how the theory is applied. In \emph{Bananaworld}\cite{Bub1} I called this interpretation ``information-theoretic,'' but here I want to emphasize the connection with Bohr's view. 

Quantum mechanics began with Heisenberg's ``Umdeutung'' paper, his proposed ``reinterpretation'' of physical quantities as noncommutative.\cite{Heisenberg} To say that the algebra of physical quantities is commutative is equivalent to saying that the idempotent elements form a Boolean algebra. For the physical quantities or observables of a quantum system represented by self-adjoint Hilbert space operators, the idempotent elements are the projection operators, with eigenvalues 0 and 1. They represent yes-no observables, or properties (for example, the property that the energy of the system lies in a certain range of values), or propositions (the proposition asserting that the value of the energy lies in this range), with the two eigenvalues corresponding to the truth values, true and false. 

Heisenberg's insight amounts to the proposal that certain phenomena in our Boolean macro-world  that defy a classical physical explanation can be explained probabilistically as a manifestation of collective behavior at a non-Boolean micro\-level. The Boolean algebra of physical properties of classical mechanics is replaced by a family of ``intertwined'' Boolean algebras, one for each set of commuting observables, to use Gleason's term.\cite{Gleason} The intertwinement precludes the possibility of embedding the whole collection into one inclusive Boolean algebra, so you can't assign truth values consistently to the propositions about observable values in all these Boolean algebras. Putting it differently, the different perspectives associated with certain Boolean algebras in the family of Boolean algebras of a quantum system, say the Boolean algebras for position and momentum, don't fit together into a single Boolean algebra, unlike the corresponding family for a classical system. Welcome to complementarity.

Bohr did not refer to Boolean algebras, but the concept is simply a precise way of codifying what I think Bohr should have had in mind by insisting (his emphasis) that\cite{Bohr1}
\begin{quote}
\emph{however far the phenomena transcend the scope of classical physical explanation, the account of all evidence must be expressed in classical terms.}
\end{quote}
 by which he meant ``unambiguous language with suitable application of the terminology of classical physics''---for the simple reason, as he put it, that we need to be able to ``tell others what we have done and what we have learned.'' Formally speaking, the significance of ``classical'' here as being able to ``tell others what we have done and what we have learned'' is that the events in question should fit together as a Boolean algebra. George Boole, who came up with the idea in the mid-1800's, introduced Boolean constraints on probability as ``conditions of possible experience.''\cite{Boole}

Bohr's primary insight was to see  that quantum mechanics is quite unlike any theory we have dealt with before in the history of physics, and so   explanation in such a post-classical theory  can't be the sort of representational explanation we are familiar with in a  theory that is commutative or  Boolean at the fundamental level. Quantum probabilities can't be understood in the classical (I would say Boolean) sense as quantifying ignorance about the pre-measurement value of an observable, but cash out in terms of what you'll find if you ``measure,'' which involves considering the outcome, at the Boolean macrolevel, of manipulating a quantum system in a certain way. 

A quantum ``measurement'' is a bit of a misnomer and not really the same sort of thing as a measurement of a physical quantity of a classical system. It involves putting a microystem, like a photon, in a situation, say a beamsplitter or an analyzing filter, where the photon is forced to make an intrinsically random transition recorded as one of  two  macroscopically distinct alternatives in a device like a photon detector. The registration of the measurement outcome at the Boolean macrolevel is crucial, because it is only with respect to a suitable structure of alternative possibilities that it makes sense to talk about an event as definitely occurring or not occurring, and this structure is a Boolean algebra.

There are various ways to see how a Boolean macroworld could emerge from the intertwined Boolean algebras at the microlevel, but they are all approximate and to some extent conceptually fuzzy. We are familiar in physics with collective phenomena at the macrolevel that can be quite different from the behavior of individual systems at the microlevel. For example, the air has a temperature, but individual air molecules don't. We understand how the temperature of the air is a collective property of large collections of air molecules, but how to explain the emergence of a commutative or Boolean macrolevel from an underlying noncommutative or non-Boolean microlevel is trickier. It is still  to some extent an ongoing research problem,\cite{Landsman} but it is not, fundamentally, an interpretative problem.\footnote{In the last section, ``The Information-Theoretic Interpretation,'' of the final chapter of \emph{Bananaworld}\cite{Bub2} I discussed Hepp's toy model of  quantum measurement, in which a macroscopic measurement pointer is represented by an infinite array of qubits, and suggested that this was a way to see how quantum probabilities of ``what you'll obtain if you measure'' could become classical probabilities in the ignorance sense of ``what's there.''  I noted Bell's critique that the transition only occurs at the infinite limit but argued that this was similar to the idealization involved in explaining phase transitions in classical physics, as in the transition from a liquid to a solid. As Matthew Leifer pointed out at a workshop on \emph{Bananaworld} at Western University, London, Ontario in June, 2016,\cite{workshop} the physical significance of a singular limit, as in Hepp's model, is questionable. Following Matt's criticism, I have completely revised this section of the chapter for a paperback edition of \emph{Bananaworld}. As I now see it, the pathology of the singular limit actually supports a Bohrian analysis of measurement in quantum mechanics. The fact that quantum probabilities only become classical ignorance probabilities at the infinite limit is another way of seeing that events at the Boolean macrolevel are outside quantum theory.}

The quantum revolution is about new sorts of probabilistic correlations in nature, analogous to the sense in which the relativistic revolution is about new sorts of spatio-temporal relations. 
What Hilbert space gives you is fundamentally a theory of probabilistic correlations. Probabilities and probabilistic correlations arise as a feature of the non-Boolean structure. They are ``uniquely given from the start,'' to quote von Neumann,\cite{vonNeumann1} related to the angles in Hilbert space, not measures over states as they are in a classical or Boolean theory.  The intertwinement of commuting and noncommuting observables in Hilbert space imposes objective pre-dynamic probabilistic constraints on correlations between events, analogous to the way in which Minkowski space-time imposes kinematic constraints on events. The probabilistic constraints encoded in the geometry of Hilbert space provide the framework for the physics of a genuinely indeterministic universe. They characterize the way probabilities fit together in a world in which there are nonlocal probabilistic correlations that violate Bell's inequality up to the Tsirelson bound, and these correlations can only occur between intrinsically random events.\cite{Bub3}

So quantum probabilities are ``sui generis,'' as von Neumann also put it.\cite{vonNeumann2} They don't quantify incomplete knowledge about an ontic state (the basic idea of ``hidden variables''), but reflect the irreducibly probabilistic relation between the non-Boolean microlevel and the Boolean macrolevel, expressed through the intrinsic randomness of events associated with the outcomes of quantum measurements. Here it is crucial, as the Frauchiger-Renner argument shows, that the reference is to  \emph{one ultimate observer system} as the end-point of a quantum mechanical analysis. Bohr repeatedly emphasizes this:\cite{Bohr2}
\begin{quote}
In the system to which the quantum mechanical formalism is applied, it is of course possible to include any intermediate auxiliary agency employed in the measuring processes. \ldots The only significant point is that in each case some ultimate measuring instruments, like the scales and clocks which determine the frame of space-time coordination---on which, in the last resort, even the definition of momentum and energy quantities rest---must always be described entirely on classical lines, and consequently be kept outside the system subject to quantum mechanical treatment. 
\end{quote}

The outcome of a measurement is an intrinsically random event at the macro\-level, not described by the deterministic dynamical evolution of the quantum theory. In this sense it is outside the theory, or ``irrational'' as Pauli characterizes it (his emphasis):\cite{Pauli}
\begin{quote}
Observation thereby takes on the character of \emph{irrational, unique actuality} with unpredictable outcome. \ldots  Contrasted with this \emph{irrational aspect} of concrete phenomena which are determined in their \emph{actuality}, there stands the \emph{rational aspect} of an abstract ordering of the \emph{possibilities} of statements by means of the mathematical concept of probability and the $\psi$-function [I would say `by means of the geometry of Hilbert space'].
\end{quote}


The ``single ultimate observer system'' constraint in the application of quantum mechanics is related to the idea of a movable ``cut'' between the observer and what's observed, and to Bohr's concept of a ``phenomenon.'' The world is fundamentally quantum (or non-Boolean), but what we want to explain are phenomena at the Boolean macrolevel. So we partition the theoretical description into a part that involves a quantum (non-Boolean) analysis and a part that involves a classical (Boolean) analysis. The cut imposes a Boolean frame on the world, implemented physically by the choice of measuring instrument. Since the choice is up to us, the cut is movable. Of course, once you decide what counts as the ultimate measuring instrument in a given scenario, the cut is no longer movable. If you move it, you are ``subdividing the phenomenon,'' as Bohr puts it, and the two analyses will be incompatible---in general they won't both fit into one Boolean frame. In effect, a phenomenon is an episode from the perspective of a Boolean frame. Information about a system provided through different Boolean frames that can't be embedded into a single Boolean frame is ``complementary'': although the information isn't additive, it provides different perspectives that would be additive and part of one Boolean frame in a classical theory.

To quote Bohr:\cite{Bohr3}
\begin{quote}
The unaccustomed features of the situation with which we are confronted in quantum theory necessitate the greatest caution as regards all questions of terminology. \ldots It is certainly far more in accordance with the structure and interpretation of the quantum mechanical symbolism, as well as with elementary physical principles, to reserve the world ``phenomenon'' for the comprehension of the effects observed under given experimental conditions. 

These conditions, which include the account of the properties and manipulation of all measuring instruments essentially concerned, constitute  in fact the only basis for the definition of the concepts by which the phenomenon is described [i.e., a Boolean frame]. It is just in this sense that phenomena defined by different concepts, corresponding to mutually exclusive experimental arrangements [incompatible Boolean frames], can unambiguously be regarded as complementary aspects of the whole obtainable evidence concerning the objects under investigation.
\end{quote}

To return to the Frauchiger-Renner argument: just one observer perspective is legitimate in the application of quantum mechanics, the perspective of the observer for whom an actual measurement outcome occurs at the macrolevel. If Alice and Bob  represent a composite ultimate observer for whom there are definite events at the macrolevel, the observer perspective is legitimate,  in which case the final state of the combined quantum coin and qubit system is  $\ket{h}_{A}\ket{0}_{B}$ or $\ket{t}_{A}\ket{0}_{B}$ or $\ket{t}_{A}\ket{1}_{B}$. If a  super-observer subsequently measures the super-observables $X, Y$ on the whole composite Alice-Bob system, the probability of obtaining the pair of outcomes $\{\mbox{ok}, \mbox{ok}\}$ is 1/4 for any of the product states  $\ket{h}_{A}\ket{0}_{B}$ or $\ket{t}_{A}\ket{0}_{B}$ or $\ket{t}_{A}\ket{1}_{B}$ (here interpreted as representing eigenstates of the super-observables $A, B$). After the measurement, the super-observables $A, B$ are indefinite, and so are the corresponding quantum coin and qubit observables. There is no contradiction  because the argument from the alternative expressions for the entangled state no longer applies. If the super-observer perspective is legitimate, there are no definite Alice and Bob events at the macrolevel and the state is the entangled state. The probability of a super-observer finding the pair of outcomes $\{\mbox{ok}, \mbox{ok}\}$ is 1/12, but there is no contradiction because there are no measurement outcome events for Alice and Bob.

 It's not that unitarity  is suppressed at a certain level of complexity, where  non-Booleanity becomes Booleanity and quantum becomes classical. Rather, there is a macrolevel, which is Boolean (Bohr would say ``classical'), and there are actual events at the macrolevel. But any system, of any complexity, is fundamentally a quantum system and can be treated as such, in principle. A unitary dynamical analysis of a measurement process goes as far as you would like it to go, to whatever level of precision is convenient.  The collapse, as a conditionalization of the quantum state, is something you put in by hand after observing the actual outcome. The physics doesn't give it to you. 

Special relativity, as a theory about the structure of space-time, provides an explanation for length contraction and time dilation through the geometry of Min\-kow\-ski space-time, but that's as far as it goes.  This explanation didn't satisfy Lorentz, who wanted a dynamical explanation in terms of forces acting on physical systems used as rods and clocks.\cite{Janssen} Quantum mechanics, as a theory about randomness and nonlocality, provides an explanation for probabilistic constraints on events through the geometry of Hilbert space, but that's as far as it goes. This explanation doesn't satisfy Everettians, who insist on a representational story about how nature pulls off the trick of producing intrinsically random events at the macrolevel, with nonlocal probabilistic correlations constrained by the Tsirelson bound. 

Do we really want to give up the concept of measurement as a procedure that provides information about the actual value of an observable of a system to preserve the ideal of representational explanation in physics? It seems far more rational to accept that if current physical theory has it right, the nature of reality, the way things are, limits the sort of explanation that a physical theory provides. 

\section*{Acknowledgements}
Although we probably don't agree, thanks to Matt Leifer for clarification of the Frauchiger-Renner argument, and to Michel Janssen and Michael Cuffaro for some really helpful input on early drafts of this paper.

\end{document}